# Excel: Automated Ledger or Analytics IDE?

Andrew Kumiega, PhD

*Assistant Professor of Analytics, Illinois Institute of Technology*

AKumiega@IIT.EDU

**ABSTRACT**

Since the inception of VisiCalc over four decades ago, spreadsheets have undergone a gradual transformation, evolving from simple ledger automation tools to the current state of Excel, which can be described as an Integrated Development Environment (IDE) for analytics. The slow evolution of Excel from an automation tool for ledgers to an IDE for analytics explains why many people have not noticed that Excel includes a fully functional database, an OLAP Engine, multiple statistical programming languages, multiple third-party software libraries, dynamic charts, and real time data connectors. The simplicity of accessing these multiple tools is a low-code framework controlled from the Excel tool that is effectively an IDE. Once we acknowledge Excel's shift from a desk top application to an IDE for analytics, the importance of establishing a comprehensive risk framework for managing this distinctive development environment becomes clear. In this paper we will explain how the current risk framework for spreadsheets needs to be expanded to manage the growing risks of using Excel as an IDE for analytics.

## 1.0 INTRODUCTION

> *"To exist is to change, to change is to mature, to mature is to go on creating oneself endlessly."*
> *Henri Bergson*

Founded in 2000, EuSpRIG emerged in response to the growing acknowledgment within professional and academic circles regarding the risks of using spreadsheets. This recognition stemmed from the understanding that while spreadsheets offered unparalleled flexibility and accessibility, it also posed significant risks because of human errors. EuSpRIG's efforts have helped increase awareness, shared knowledge, and created best practices to mitigate errors in spreadsheets. The organization has influenced how professionals in various sectors perceive and address spreadsheet risk. Along with its focus on spreadsheet risk, EuSpRIG has also played a role in introducing spreadsheet users to new productivity tools that are continuously being added into Excel (Kemmis and Thomas 2008; Paine 2008). As even the accounting profession recognizes the shift from reporting to data analysis, the urgency for Microsoft to transform Excel from an accounting tool to a data analysis platform is growing (Jackson, Michelson, and Munir 2023). The question under discussion is whether Excel is still an automated bookkeeping tool or whether Excel has been transformed into an IDE for analytics applications.

## 2.0 WHAT IS THE PROGRAMMING LANGUAGE OF WHICH EXCEL IS A PART?

There are three major programming paradigms: procedural, object-oriented, and functional. Procedural, exemplified by C or Pascal, is step-based. The next paradigm is object-oriented, which focuses on objects. Examples of this paradigm are C++ and Java. The final paradigm is functional, which treats computation as mathematical functions and emphasizes immutability. Modern interactive development environments (IDE), such as Visual Studio, allow programmers to mix and match these paradigms. The Excel IDE also currently has development capabilities in all three paradigms.

Excel's primary programming structure should be considered functional since most Excel users create immutable complex functions for each individual cell. Panko's seminal 2000 paper within EuSpRIG clarifies the distinctive risks associated with empowering users to use functions in Excel for data manipulation (R. Panko 2000). The paper identifies and emphasizes four critical types of errors that have since become foundational in the discourse of EuSpRIG. These errors include quantitative and qualitative errors, omission errors, logic errors, and mechanical errors. Panko's work serves as a cornerstone for understanding and addressing these pitfalls, mainly focusing on devising strategies to prevent these errors. The original EuSpRIG papers, inspired by Panko's insights, laid the groundwork for best practices in mitigating risks associated with user-driven data manipulation in Excel.

Spreadsheet errors were the focus of EuSpRIG's early research (Ayalew, Clermont, and Mittermeir 2008; 2008; R. R. Panko 2008; Rajalingham, Chadwick, and Knight 2008). The original body of work focused on auditing and error prevention. The organization's website lists 233 papers from its annual conferences, covering categories like computers and society, software engineering, access databases, human-computer interaction, programming languages, presentation, and others. That said, many of the most recent papers focus on the risks posed by the new productivity tools being added to Excel. Examples include: "Should Power Query Tools Change Spreadsheet Risk Approaches", "Mitigating Spreadsheet Model Risk with Python Open Source Infrastructure", "Git and Excel", and "Structured Spreadsheet Modeling and Implementation with Multiple Dimensions" (Bartholomew, n.d.; Hurst, n.d.; Mireault 2019). These papers mainly focus on an individual productivity tool risk but do not acknowledge that Excel has evolved into an IDE for analytics where users regularly create applications using all these low-code tools.

## 3.0 EXCEL HAS EVOLVED INTO AN IDE

### 3.1 Excel Transformation

Excel's transformation over the last three decades from a simple ledger to a development environment for high level analytics, including machine learning, has been gradual. The magnitude of this shift has gone mostly unnoticed by many users who still use Excel to build traditional applications. However, Microsoft has embedded entire new development platforms into Excel, as described in Table 1. Each addition introduces more software risks, requiring the adoption of new development methodologies and testing techniques. The inclusion of multiple tools in a spreadsheet has exponentially increased the complexity of the final application.

**Table 1.**

| Sub-Section | Tool | Tool functions | |
|---|---|---|---|
| Data management | Power Query | Get and Transform Data | Enables connections to databases, text files, workbooks, and online services. |
| Data management | Power Query | Query Editor | Interfaces transforming and cleaning data. |
| Data management | Power Query | Power Query Formula Language | The M Language is a functional language for both Excel and PowerBi. IT allows |

| | | | for complex manipulation of data in Power Query. |
|---|---|---|---|
| Data Management and Data Shaping | Power Pivot Tab | Diagram View | Lets users view, create, and change relationships in a graphical format, making it easier to understand the structure of the data model. |
| Data Shaping | Advanced Filter | Filtering | Advanced Filter: This tool offers more sophisticated filtering options, letting users specify complex criteria for extracting data from a data set. |
| Data Shaping | Advanced Filter | Slicing | Slicing allows analysts to perform analysis on small portions of the data. |
| Data Shaping/Business Analytics | Power Pivot Tab | Power Pivot | Power Pivot is an OLAP style tool that allows analysts to quickly summarize data. Power Pivot also allows the analysts to create pivot graphs that allows management to quickly understand the data. |
| Business Analytics and Intelligence | Power Map Tab | Power Map | Power Maps, enable users to plot data points on maps, visualize trends, patterns, and relationships, and gain insights into geographic data sets. |
| Business Analytics And Intelligence | Data Tab | Data Analysis ToolPak | This is the original data analysis tool. It provides a range of statistical and analytical tools to perform data analysis tasks such as regression analysis, histograms, and more. |
| Business Analytics And Intelligence | Data Tab | Solver | Solver lets an end user perform linear and non-linear optimization. |
| Business Analytics And Intelligence | Data Tab | Advanced Solver/XLMiner | XLMiner helps with various data analysis and predictive modeling tasks, including regression analysis, classification, clustering, decision trees, random forests, and forecasting. |
| Business Analytics And Intelligence | SAS Add-in | SAS Tab | Allows for the full functionality of SAS to be implemented within Excel. |
| Programming | Developer Tab | Visual Basic for Applications (VBA) | Excel VBA enables users to automate Excel processes and |

| | | | create intricate custom calculations. |
|---|---|---|---|
| Programming | Python | Python | Excel Python enables advanced data analysis, visualization, and automation using Python libraries within the Excel environment by integrating Python directly into Excel. |

Incorporating all these programming tools into Excel has effectively created an IDE for self-service analytics professionals. Management of these tools is facilitated by additional low-code interfaces that are conveniently accessible from within the toolbar. Many of these interfaces effectively act as a custom IDE for these analytic tools. The Power Query toolbar is effectively SQL Server's graphical query designer. Power Pivot is effectively a graphical designer for OLAP. Excel offers seamless integration with various low-code analytics platforms like the Advanced Solver (XLMiner) Suite, XLStat Premium, SAS, and others through toolbar ribbons. The recent inclusion of Python in Excel further enhances users' ability to develop sophisticated analytics applications. Most of these analytics tools including Python further the low-code framework by including AutoML functionality.

The low-code interfaces empower users to create intricate analytics solutions with unprecedented ease and efficiency, eliminating the necessity for advanced programming expertise. When considering these advancements collectively, it becomes evident that Excel is gradually metamorphosing into an IDE for analytics. Consequently, the forefront of risk in Excel is swiftly transitioning from conventional risks to those associated with programming for complex analytics applications.

### 3.2 More Automation Equals More Risk

As Excel evolves from a simple spreadsheet to a powerful analytics IDE that can access and manage a variety of built-in analytics tools, it becomes clear that EuSpRIG's expertise must evolve accordingly. For each of the low-code tools or programming tools rapidly being installed into the Excel IDE a new best practice for that tool needs to be developed. As self-service analysts move from using individual spreadsheets to deploying analytic applications online, EuSpRIG must evolve its best practices to effectively manage the software architecture of these distributed analytic applications. EuSpRIG will need to add at least four new research areas: Data Management, Data Shaping, Business Analytics/Business Intelligence, and Analytic Software Architecture

### 3.3 Data Management: Opportunities and Risks

As shown in Table 1 several tools in Excel will let users import, manipulate, and analyze large data sets within Excel. It is estimated that Data Scientists spend a majority of their time gathering and organizing data sets. Excel's data automation tools enable analysts to efficiently build a data warehouse directly in Excel. With Power Query, analysts can seamlessly import and merge data from multiple sources-databases, text files, websites, and other Excel files-without the need for complex SQL queries. In addition, Power Query provides a set of low-code tools for cleansing, transforming, and manipulating data before integration with Excel. In addition, similar to Excel's record macro functionality, Power Query generates M-code at the end of the data shaping process, allowing analysts to modify it directly for more complex data shaping. This entire process of data import and transformation is performed by the analyst, freeing central IT resources to focus on enterprise-wide data management initiatives.

By reducing an analyst's dependency on central IT, they can tailor data manipulation processes according to their unique preferences and business context, leading to more accurate and relevant insights. Users can quickly adapt to changing business needs and explore new data sources or analysis techniques without extensive IT intervention. Yet this increase in productivity comes with more risks. Most analysts are not trained in database design or in data gathering techniques. The lack of database skills and Excel's new Get Data functionality leads to new risks such as data corruption from unstable data sources such as free web pages that regularly change the data structure.

The risks associated with building a data warehouse in Excel are similar to those in a SQL Server environment. Poorly designed queries can overwhelm hard drives and CPUs, potentially leading to system crashes. In addition, combining data into a single file in Excel or Python via a low-code environment creates risks because errors made by untrained users can jeopardize data integrity. These errors, which occur during the import and consolidation process, can include inadvertently deleting or changing data.

Most Excel users are not trained to manage a production data warehouse. So users must be trained to manage a production data warehouse. EuSpRIG has historically addressed these concerns within its body of knowledge, but the size and complexity of Excel based data warehouses has gone up significantly over the last five years (Murphy, 2008). As the use of Power Query expands, EuSpRIG will probably introduce many new best practices for creating internal data warehouses in Excel (Chadwick 2008).

The following data governance tasks are essential for Excel based data warehouses:

- Documentation: Users need to document each step of the process, state its purpose, and define a test for it. Stored scripts should undergo rigorous code review and quality testing, similar to production SQL statements.

- Data source mapping: To maintain data integrity, it is critical to map each data warehouse to its original data source within the organization.

- Retesting: To ensure continued accuracy and reliability, any changes to the original database structure require retesting of Power Query steps. Before any user of these Excel data warehouses uses the warehouse as input for a new analytic application, the integrity of the data should be retested.

- Annual Audit: Annual auditing of all spreadsheets with embedded warehouses is critical to ensure imported data matches the original column and data structures.

### 3.4 Data Shaping: Opportunities and Risks

Data shaping involves refining and structuring raw data to facilitate effective analysis and reporting. Data shaping includes tasks such as cleaning, filtering, aggregating, and transforming data sets to reveal meaningful insights. Advanced data shaping capabilities are available in both Power Query and Power Pivot. These productivity tools are essential for analysts to effectively organize data for both descriptive and predictive analysis. With Power Query, analysts can streamline tasks such as eliminating duplicates, filtering rows based on criteria, splitting columns, and generating custom data using formulas. Meanwhile, PowerPivot facilitates the efficient transformation of raw data in the data warehouse into summary tables. These summary tables serve as the basis for creating tables and pivot charts that management relies on to identify trends, patterns, and anomalies in the data. The summary tables are also regularly fed into analytical algorithms, such as regression and gradient boosted trees, to provide management with predictive insights.

While data shaping tools enhance analysts' productivity, there are inherent risks in allowing them to shape data without best practices. Analysts may lack understanding of the problem or data set, leading to improper filtering, transformation, or cleansing of the data set (O'Beirne 2008). Users are also vulnerable to formula errors, including syntax or logic errors (Grossman and Ozluk 2010). The addition of this functionality to the Excel IDE shifts data risk focusing on cell-to-cell data dependencies to what is normally a SQL database risk.

Existing data risk management best practices outlined in the EuSpRIG literature need to evolve to include not only the known data risk traditional spreadsheet data risks, but also risks associated with data design and data warehouse management. (Mittermeir, Clermont, and Hodnigg 2008). It is doubtful that these data warehouses will be managed through the data governance structure of the enterprise. It is therefore reasonable to assume that analysts will re-use and share these data structures regularly to create other analytical solutions. To address the risks linked with new data shaping and management tools in Excel, EuSpRIG must expand its existing best practices for managing data, as outlined above. To ensure accurate results, in addition to the data governance measures discussed earlier, specific data governance procedures must be established for the data shaping phase. At a minimum these controls should focus upon:

1. **Data Transformation Consistency**:

    - In analytics, many data transformations can be easily applied to data sets using Excel's IDE without the need for coding. Standard transformations like mean, median, percentile, and z-score, among others are all standard functions in Excel, power pivot, and Power Query. In addition, with Python embedded in Excel, it has become easy to create custom algorithms for removing outliers. Since each transformation can result in a different predictive algorithm, it is important to build models with alternative transformations. The data governance process for these transformations should include the rationale for using the transformation along with how the transformation changed the accuracy of the algorithm.

    - Ideally, there should be consistency in the transformations between data items.

    - Standardize data shaping procedures and enforce adherence to established standards and best practices across the organization.

2. **Auditing of Data Shaping:**

    - The data shaping step should follow traditional risk auditing criteria which is once a year or if there is any change in the process. This is especially important in regulated industries such as finance or insurance to ensure the data meets regulatory requirements.

3. **Management of Outliers:**

    - The process that manages outliers needs to be documented and vetted. Ideally, the analytics will be calculated with the outliers included and not included to determine the Black Swan risk for the algorithm.

### 3.5 Business Analytics and Intelligence: Opportunities and Risks:

The Excel IDE has recently been enhanced with the integration of a variety of advanced analytical tools. The tools include Power Maps, Power BI, SAS, and XLMiner (Advanced Solver). These tools feature easy-to-use, low-code development interfaces. They let analysts build complex business analysis dashboards and develop predictive algorithms directly in Excel. They also make it easy to quickly create sophisticated dashboards that can be easily distributed using Excel for the web, a cloud-based web browser-accessible version of Excel. Along with seamless collaboration, this platform offers real time capturing and seamless integration with One Drive for productivity and accessibility.

While the combination of these tools into a single low-code IDE structure heralds a significant potential for the generation of valuable business insights, it's important to recognize that, like any flexible programming environment, it can expose organizations to risk. A new key risk for these sets of tools is model risk since analysts can and do create models that are un-explainable due to the model complexity. Also, many of these tools are adopting an AutoML structure where the tool selects the best model so the analysts may not even understand the logic or the mathematics behind the selected model.

Therefore, at a minimum the concept of Explainable Artificial Intelligence (XAI) should be included in the development stage of the model. At a minimum the following XAI techniques are required to ensure that the analytical model is understandable.

1. **Model Comparison and Selection:**

    - Develop multiple models, including regressions, decision trees, linear models, and rule-based systems. Compare and rank these models to help analysts understand why the AutoML tool selected a particular model. In addition, when model rankings are close, prioritize the selection of the simplest model to improve interpretability and ease of implementation.

2. **Feature Importance Analysis**:

    - Conduct feature importance analysis to identify the factors that contribute most significantly to the model's predictions. This helps stakeholders understand which features are driving the model's decisions.

3. **Model Documentation:**

    - Create detailed documentation for each predictive algorithm, including details such as training data, hyperparameters, cross-validation techniques, and evaluation metrics. This thorough documentation allows both the analyst and the end user to choose the right model.

4. **Regulation Testing**:

    - The final model's features should be tested to determine whether they are able to predict illegal features such as age, sex, and race. This testing needs to be fully documented.

### 3.6 Software Architecture: Opportunities and Risks:

Excel has long provided an IDE for VBA programming. This IDE provides a blend of low-code and traditional coding capabilities. In the low-code environment, users can automate tasks using the recorder, while developing complex applications in the traditional coding environment. To mitigate risk, EuSpRIG has promoted a robust programming methodology for building VBA applications (Kumiega and Van Vliet 2008).

Excel's recent announcement of adding Python as a programming language has both benefits and challenges. On the one hand, it mitigates a significant risk in developing analytical applications using Python. This is because Microsoft normally ensures backward compatibility for their supported languages, which will remove a known risk with open source software. In any event, Python will introduce new risks because embedding custom Python functions into the cells will make auditing a spreadsheet complex. Assuming the Excel IDE will support the development of Python functions like VBA functions then the spreadsheet audit will need to be expanded into a full code review for model validation and verification.

Despite the final architecture between Python and Excel, it's clear that new techniques for managing both the software audit risk and the function calculations used in these hybrid Excel applications that use Python will become more complex. So the current EuSpRIG best practices for spreadsheet auditing will also need to be revisited to manage the risk from using multiple programming paradigms in a single application. The best practices will need to include traditional software testing techniques such as code reviews and test cases for cell functions that use Python libraries (Kumiega and Van Vliet 2011).

### 4.0 CONCLUSION:

Many of the historical EuSpRIG presentations have focused on new tools to increase the productivity of Excel, so the productivity improvements are already part of the EuSpRIG body of knowledge (Birch, Lyford-Smith, and Guo 2018; Grossman, Ozluk, and Gustavson 2009; Paine 2008). The evolution of Excel's functionality has been profound in recent years, with the integration of powerful features like Power Query, Power BI, Power Pivot, and Power Maps by Microsoft. The addition of these analytic tools has shifted Excel to a self-service analytics platform. Incorporating Python, SAS, XLMiner (Advanced Solver) and many other analytic add-ins into Excel has effectively transformed it into a low-code IDE for analytics. While these enhancements significantly expand Excel's capabilities, they also introduce new risks associated with data analysis and manipulation.

EuSpRIG's purpose "to promote the free exchange of information on the risks associated with the use of electronic spreadsheet programs, methods for developing and testing business models and applications created using such programs, and tools and techniques for auditing such business models and applications" is even more important today due to the expansion of Excel into an analytics IDE. This expansion brings EuSpRIG into the realm of analytic modeling via Excel. This is critical as the demand for skilled analysts using such tools is expected to grow by 35 percent between 2022 and 2032 ("Data Scientists : Occupational Outlook Handbook: : US Bureau of Labor Statistics," n.d.). With the growth of Excel through low-code tools, there's an urgent need for an organization like EuSpRIG to guide the research and development for the best practices to mitigate these new risk vectors.